\documentclass[aip,rsi,reprint,amsmath,amssymb]{revtex4-2}

\usepackage{graphicx}
\usepackage{bm}
\usepackage{siunitx}
\usepackage{booktabs}
\usepackage{microtype}
\usepackage{xcolor}
\usepackage{hyperref}
\usepackage{listings}
\usepackage{enumitem}
\usepackage{comment}

\hypersetup{
  colorlinks=true,
  linkcolor=blue,
  citecolor=blue,
  urlcolor=blue
}

\newcommand{\fa}{f_a}
\newcommand{\Pa}{P_a}

\begin{document}

\title{Spectral shaping of Gaussian white noise for synthetic axion signal generation in microwave cavity haloscopes}

\author{D.~Vattolo}
\email[Corresponding author: ]{davide.vattolo@studenti.unipd.it}
\affiliation{Dipartimento di Fisica e Astronomia, Padova, Italy}

\author{D.~Ahn} \affiliation{INFN, Sezione di Padova, Padova, Italy} 

\author{G.~Carugno} \affiliation{INFN, Sezione di Padova, Padova, Italy} 

\author{R.~Di Vora}\affiliation{INFN, Laboratori Nazionali di Legnaro, Legnaro, Padova, Italy}

\author{D. Maiello}\affiliation{Dipartimento di Fisica e Astronomia, Padova, Italy}\affiliation{INFN, Sezione di Padova, Padova, Italy} 

\author{A.~Ortolan} \affiliation{INFN, Laboratori Nazionali di Legnaro, Legnaro, Padova, Italy}

\author{G.~Ruoso}\affiliation{INFN, Laboratori Nazionali di Legnaro, Legnaro, Padova, Italy}

\author{G.~Sardo Infirri}\affiliation{Dipartimento di Fisica e Astronomia, Padova, Italy}\affiliation{INFN, Sezione di Padova, Padova, Italy} 

\author{C.~Braggio}\email[Corresponding author: ]{caterina.braggio@unipd.it}
\affiliation{Dipartimento di Fisica e Astronomia, Padova, Italy}
\affiliation{INFN, Sezione di Padova, Padova, Italy} 

\date{\today}

\begin{abstract}
Microwave cavity haloscopes search for dark-matter axions through the weak electromagnetic field generated by axion--photon conversion in a static magnetic field. Realistic synthetic signals are essential for end-to-end calibration of the receiver, validation of the analysis pipeline, measurement of signal
recovery efficiency, and blind-injection studies. Here, we present a general method for producing stochastic synthetic axion waveforms by shaping Gaussian white noise in the frequency domain. The target axion power spectral density is imposed directly on the Fourier coefficients, and arbitrarily long continuous time streams are generated efficiently using an overlap-save implementation.
The method reproduces both the ensemble-averaged spectral line shape and the time-domain fluctuations expected for a classical axion field, while providing direct control of the total injected power. Because the spectral model enters only through a replaceable transfer function, the routine can accommodate the
Standard Halo Model as well as nonstandard velocity distributions and narrow substructure. 
\end{abstract}

\maketitle

\section{Introduction}

The Lagrangian of quantum chromodynamics (QCD)  contains a term that violates CP symmetry\cite{TASIhook:2023}:
\begin{equation}
\mathcal{L}_{\bar{\theta}}
=
\bar{\theta}\,
\frac{g_s^2}{32\pi^2}
G_{\mu\nu}^{a}\widetilde{G}^{a\mu\nu},
\label{eq:theta-term}
\end{equation}
where $\bar{\theta}$ is the physical strong-CP phase, $G_{\mu\nu}^{a}$ is the gluon field-strength tensor, and $\widetilde{G}^{a\mu\nu}$ is its dual.
Nonperturbative QCD methods relate a nonzero $\bar{\theta}$ to hadronic CP-violating observables, most notably the neutron electric dipole moment. 
The current experimental bound~\cite{Abel:2020aa}, $|d_n|<1.8\times10^{-26}\,e\,\mathrm{cm}$, constrains the strong-CP phase to $|\bar{\theta}|\lesssim10^{-10}$, whereas no symmetry of the Standard Model requires such a small value. This apparent
fine tuning is the strong-CP problem.

The Peccei--Quinn mechanism replaces the constant angle $\bar{\theta}$ with a dynamical field whose potential is generated by QCD and whose minimum lies at the CP-conserving value. The associated pseudo-Nambu--Goldstone boson is the axion\cite{Peccei:77,Peccei:1977aa,Weinberg:1978aa}. In addition to solving the strong-CP problem, the QCD axion is a well motivated, cold dark matter (DM) candidate: nonthermal production mechanisms can generate a cosmologically long-lived population with negligible primordial velocity dispersion \cite{Di-Luzio:2020aa}.

A leading strategy for detecting axion dark matter is the microwave cavity haloscope proposed by Sikivie \cite{Sikivie:1983aa}, which exploits the axion--photon interaction. While the strong-CP problem originates from the gluonic operator $G_{\mu\nu}^{a}\widetilde{G}^{a\mu\nu}$, experimental axion searches with haloscopes
probe the coupling of the axion field to the analogous electromagnetic
operator $F_{\mu\nu}\widetilde{F}^{\mu\nu}$, corresponding to an interaction
proportional to $a\,\mathbf{E}\cdot\mathbf{B}$:
\begin{equation}
\mathcal{L}_{a\gamma\gamma}
=
-\frac{1}{4}g_{a\gamma\gamma}a
F_{\mu\nu}\widetilde{F}^{\mu\nu}
=
g_{a\gamma\gamma}a\,\mathbf{E}\cdot\mathbf{B},
\end{equation}
This interaction enables the conversion of axions into photons in the
presence of a strong external magnetic field.
The local DM axion field can thus convert into a weak electromagnetic excitation which is resonantly enhanced when the process takes place inside a microwave cavity with a resonant mode tuned to the axion Compton frequency $\fa \simeq m_a c^2/h$.

For the local dark matter density $\rho_{\rm DM}\simeq 0.3$--$0.45~\mathrm{GeV\,cm^{-3}}$, the combination of the large axion number density and their nonrelativistic velocity dispersion results in a very large phase-space occupation number, allowing the Galactic axion population to be treated as a classical stochastic field. 

In addition, within the Standard Halo Model, axion velocities are
approximately Maxwell--Boltzmann distributed in the Galactic frame.
The Galactic escape velocity and the motion of the laboratory modify the corresponding laboratory-frame speed distribution and hence the expected axion spectral line shape~\cite{Gramolin:2022aa}. The kinetic-energy distribution maps into frequency through

\begin{equation}
f = \fa\left(1+\frac{v^2}{2c^2}\right),
\label{eq:velocity-frequency-map}
\end{equation}
so that the fractional linewidth is of order $v^2/c^2\sim 10^{-6}$. 

From an observational perspective, these properties translate into a signal that is not a deterministic monochromatic tone, but rather a narrow-band stochastic process whose amplitude and phase fluctuate over the characteristic axion coherence time.

The expected conversion power is extremely small, typically in the
$10^{-24}$--$10^{-22}\,\mathrm{W}$ range depending on the apparatus and axion model. Consequently, a haloscope result depends on an accurate characterization of the complete measurement chain, including the cavity response, microwave
components, amplifiers, digitizer, spectral estimation, filtering, and candidate-selection procedure. Synthetic axion signals provide a direct end-to-end test of this response. They can be used to verify signal recovery, measure analysis efficiency, calibrate injected power, study systematic effects, and perform blind injections.

Existing implementations have generated synthetic axion-like spectra through sequences of frequency-hopped tones. Here we present a complementary, fully stochastic construction based on frequency-domain shaping of Gaussian white noise. The method directly produces a continuous time series with the chosen ensemble power spectral density, preserves the random time-domain
behavior associated with finite coherence, and scales efficiently to long records through blockwise fast Fourier transforms (FFTs). The spectral model is modular, so alternative halo distributions or cold substructure can be included without changing the core algorithm.

\section{Method}

\subsection{Design requirements}

A useful synthetic axion generator should reproduce the target
ensemble-averaged spectral line shape, retain the stochastic amplitude and phase fluctuations of the axion field, provide direct control of the total signal power, generate records much longer than the coherence time without discontinuities and accommodate arbitrary, potentially time-dependent axion spectral
distributions, including nonstandard halo models and velocity
substructures such as axion streams~\cite{OHare:2026aa,Gramolin:2022aa}.
From a computational perspective, the method should remain efficient for the generation of long time-domain records, with manageable memory requirements, while avoiding distortions arising from the discrete nature of the digitized signal. The Nyquist theorem limits the faithfully representable frequency range to $\left[-f_s/2,+f_s/2\right]$~\cite{Shannon:49}, where the sampling rate $f_s$ is mainly limited by the available computational resources and the sampling capabilities of the digital-to-analog converter (DAC) device.

The present method meets these requirements by treating the desired axion
waveform at low frequency, with $f_a^{\prime}< f_s/2$, as the output of a linear filter driven by zero-mean Gaussian white
noise. If $x[n]$ denotes the input sequence and $h[n]$ the response of
the filter designed to implement correlations, the synthetic signal is
\begin{equation}
s[n]=(h*x)[n].
\label{eq:time-domain-filter}
\end{equation}

The requirement for a high-frequency axion signal is therefore achieved by up-converting the low-frequency signal through a mixer such that $f_a=f_a^{\prime}\pm f_{\rm LO}$, with $f_{\rm LO}$ is the local oscillator frequency. 
Since the proposed method generates replicas of the original spectrum at integer multiples of $f_s$, experimental tests are also required to verify that their amplitude is sufficiently suppressed at $f_a$.

\subsection{Target axion spectrum}
In haloscopes, particularly in the context of axion-photon coupling, the detected signal $s(t)$ is proportional to the axion field.
By Parseval’s theorem, the spectral lineshape can be related to the Fourier transform $\widetilde{S}(f)=|\tilde{s}(f)|^2$ of the signal, such that
\begin{equation}
\frac{1}{T}\int_0^T |s(t)|^2\,dt
=
\int_{-f_s/2}^{f_s/2}|\tilde{S}(f)|^2\,df
=
P_a.
\label{eq:psd-normalization}
\end{equation}
where $\Pa$ is the desired mean signal power. 
For a generic laboratory-frame axion speed distribution
$p_{\rm lab}(v)$, the corresponding one-sided spectral line shape is
obtained by mapping the axion kinetic energy into frequency,
\begin{equation}\label{eq:lineshape}
S_a(f)
= P_a\,p_{\rm lab}\!\left[v(f)\right]
\left|\frac{dv}{df}\right|,
\end{equation}
where
\begin{equation}
v(f)=c\sqrt{2\left(\frac{f}{f_a}-1\right)},
\end{equation} 
as follows from eq.\,\ref{eq:velocity-frequency-map}.

For the laboratory-frame Maxwellian distribution adopted in this work, the resulting spectral line shape can be conveniently parametrized in terms of two quantities, $\alpha$ and $\beta$. The parameter $\alpha=v_e/v_0$ accounts for the boost of the laboratory with respect to the Galactic halo, while $\beta$ sets the characteristic frequency scale of the spectrum. In particular,
\begin{equation}
\beta =
\left(\frac{c}{\bar{v}}\right)^2\frac{1}{f_a},
\label{eq:beta}
\end{equation}
so that the characteristic spectral width is
$\Delta f_a =1/{\beta}$.
For the reference values $\bar{v}=270~\mathrm{km\,s^{-1}}$ and
$v_e=220~\mathrm{km\,s^{-1}}$, one obtains $\alpha\simeq0.815$ and
$\beta\simeq1.2\times10^6/f_a$.

With this parametrization, for $f\geq f_a$ the laboratory-frame spectral line shape is
\begin{equation}
S_a(f)
=P_a
\frac{\beta}{\alpha\sqrt{\pi}}\,
e^{
-\beta\left(
\frac{\alpha^2}{\beta}+f-f_a
\right)}
\sinh\left(
2\alpha\sqrt{\beta(f-f_a)}
\right),
\label{eq:axion-lineshape}
\end{equation}
 and $S_a(f)=0$ for $f<f_a$.

In the limiting case of an isotropic, unboosted Maxwell--Boltzmann
distribution, $\alpha\rightarrow0$, Eq.~\ref{eq:axion-lineshape}
reduces to
\begin{equation}
S_a(f)
=P_a
\frac{2\beta^{3/2}}{\sqrt{\pi}}
\sqrt{f-f_a}\,
e^{-\beta(f-f_a)}
\label{eq:unboosted-lineshape}
\end{equation}
for $f\geq f_a$.
A two-sided spectrum for a real waveform is obtained by assigning one half of the total power to each of the peaks at $\pm\fa$.

Note that, since the target spectrum enters the algorithm only through $S_a(f)$ in Eq.\,\ref{eq:lineshape}, alternative velocity distributions can be implemented without modifying the signal-generation procedure, including cold streams and fine-grained dark-matter substructure that may produce narrow spectral features in axion haloscopes~\cite{OHare:2018aa,OHare:2026aa,Yi:2023aa}.

The numerical spectrum is finally discretized on the FFT frequency grid and renormalized so that its Riemann sum equals $\Pa$.

\subsection{Spectral shaping of Gaussian white noise}

Since the axion signal $s(t)$ is a stationary stochastic process, it is fully characterized by its autocorrelation function $R(\tau)=\mathbb{E}[s(t)s(t+\tau)]$, related to the power spectral density (PSD) via the Wiener--Khinchin theorem, $S_a(f)=\mathcal{F}{R(\tau)}$~\cite{Papoulis:1967,Wiener:30}.

We generate the synthetic axion signal as the output of a linear time-invariant (LTI) filter driven by white noise~\cite{Oppenheim:1999}. A discrete Gaussian uncorrelated sequence $x[n]\sim\mathcal{N}(0,\sigma^2)$ is generated at sampling frequency $f_s$ via a seeded pseudorandom generator, with $\sigma^2=f_s$ ensuring a normalized PSD. 

Filtering $x[n]$ with impulse response $h[k]$ chosen to reproduce the axion autocorrelation, and using $R_{xx}[m]=\sigma^2\delta[m]$, gives $R_{sx}[m]=\sigma^2 R_{ss}[m]$~\cite{Oppenheim:1999}. By the convolution theorem, this is equivalent in the continuous frequency domain to shaping the flat input PSD by the target spectral line shape,
\begin{equation}
S(f)=|\widetilde{h}(f)|^2,
\label{eq:axion_PSD}
\end{equation}
consistent with the Wiener--Khinchin relation for the shaped output.

In practice, the discrete PSD is estimated via the average of $M$ periodograms~\cite{Vaseghi:1996},
\begin{equation}
S_M(f_k)=\frac{1}{M}\sum_i^M|\tilde{s_i}(k)|^2 \quad \xrightarrow[M\rightarrow \infty] \,\,\,\, S(f_k)
\label{eq:discrete-PSD}
\end{equation}
where $f_k={k\Delta f}/{N}$ are the discrete frequencies with frequency resolution $\Delta f=f_s/N$ corresponding to $N$ FFT samples. Here $\tilde{s_i}(k)$ is the discrete Fourier transform coefficient of the $i$-th periodogram calculated from the  discrete time series s[n].

The synthetic axion waveform is obtained via inverse FFT of $S[k]$, enforcing Hermitian symmetry (or using a real-input FFT) to guarantee $s[n]\in\mathbb{R}$. The procedure applies equally at intermediate frequency or in complex baseband with independent I/Q components.

\subsection{Continuous generation by overlap-save processing}
In principle, the entire synthetic waveform could be generated at once by producing all $N_{\rm tot}$ white-noise samples, computing their Fourier transform, applying the spectral filter $\widetilde{h}(f_k)$, and transforming the result back to the time domain. Although straightforward, this approach becomes impractical for very long waveforms, since both the computational cost and, more importantly, the memory requirements increase with the total number of samples $N_{\rm tot}$. In particular, the complete time series and the corresponding complex Fourier-domain arrays must be simultaneously stored in memory. To overcome this limitation, we generate the signal blockwise using fixed-size FFTs.

Importantly, the axion filter has a finite temporal memory, with a
characteristic timescale determined by the axion spectral linewidth, $\tau_a \sim 1/\Delta f_a$. Consequently, the signal at a given time depends not only on the current white-noise sample, but also on a finite history of preceding samples. Simply filtering consecutive blocks
independently would artificially reset this memory at each block boundary, thereby breaking the temporal correlations of the synthetic axion field. We therefore implement the frequency-domain filtering using the overlap-save method~\cite{Proakis:1996:OLS}, which reproduces the linear convolution of a continuous input stream using fixed-size FFT blocks.

The need for overlap-save also follows from the distinction between linear and circular convolution. Multiplication in the Fourier domain corresponds to circular convolution, whose output has the same length $N$ as the DFT and implicitly assumes periodicity with period $N$. In contrast, the linear convolution of two sequences of length $N$ has length $2N-1$. The samples extending beyond the $N$-sample interval are therefore wrapped around to the beginning, introducing boundary artifacts that can distort the synthesized waveform. The overlap-save method eliminates these corrupted samples and retains only the portion corresponding to the desired linear convolution.

Let the finite impulse response associated with $\widetilde{h}(f_k)$ have an effective length $M$, and let each FFT block contain $N$ samples, with $N\geq M$. The overlap-save algorithm proceeds as follows:

\begin{enumerate}[itemsep=0pt]
\item construct an input block containing the last $M-1$ samples from the
previous white-noise block, or $M-1$ zeros for the first block, followed by $L$ newly generated samples;
\item compute its $N$-point FFT;
\item multiply the result by the precomputed transfer function $\widetilde{h}(f_k)$
\item compute the inverse FFT;
\item discard the first output samples $M-1$, which are affected by the circular-convolution wrap-around;
convolution, and retain the remaining $L$ samples;
\item repeat and concatenate the retained samples.
\end{enumerate}

The procedure therefore yields $L=N-M+1$ valid output samples per iteration.

The filter length must be chosen long enough to represent the correlation time of the narrow-band signal. In practice, convergence is checked by increasing $M$ and verifying that the recovered spectrum and autocorrelation are unchanged within the required tolerance. The FFT size can be selected independently for computational efficiency, commonly as a power of two.

The computational cost scales approximately as $\mathcal{O}(N_{\mathrm{tot}}\log N)$ for a record containing $N_{\mathrm{tot}}$ samples, while the memory requirement is set by a single FFT block. 

\section{Validation of the generated waveform}

\subsection{Time-domain properties}

\begin{figure}[htp]
\begin{center}
\includegraphics[width=1\linewidth]{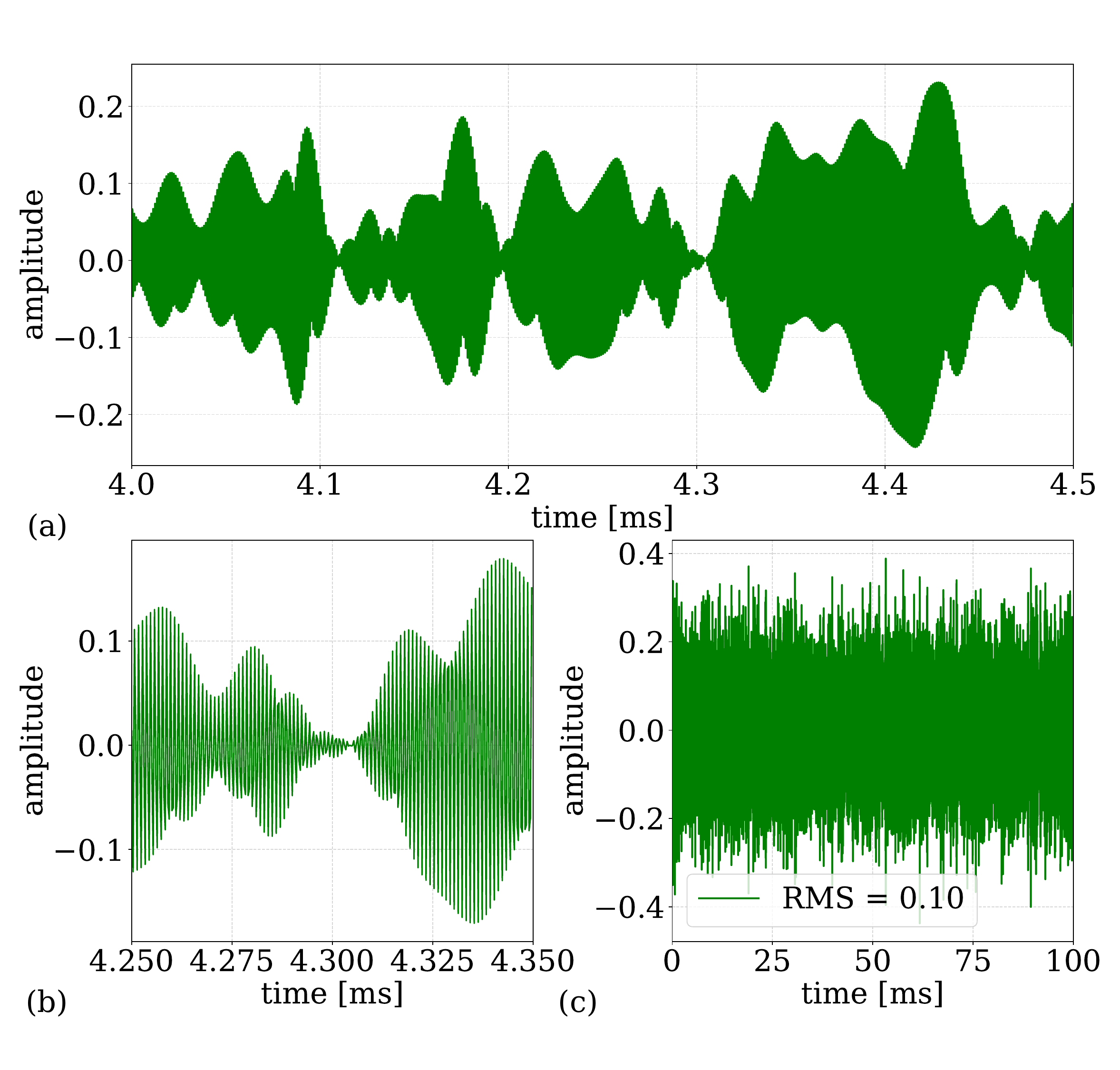}\caption{Synthetic axion signal generated considering a 6~GHz axion, so with $\beta = 2.058 \times10^{-4}~\mathrm{Hz}^{-1}$, but with oscillating frequency set at $f_a = 1~\mathrm{MHz}$ for convenience. The power was set to $P_a =0.01~\mathrm{V}^2$ and the signal was sampled at $f_s = 3~\mathrm{MHz}$.
(a) Alternating intervals of enhanced and suppressed amplitude resulting from constructive and destructive interference among the spectral components of the axion field. (b) Zoomed-in view showing the carrier oscillation at $f_a=1~\mathrm{MHz}$. (c) Over a longer time interval, the waveform exhibits the noise-like behavior characteristic of the stochastic axion field. The RMS amplitude is determined by the set signal power according to $P_a = |\mathrm{RMS}|^2$.}
\label{fig:2}
\end{center}
\end{figure}

A representative realization of the generated synthetic axion signal is shown in Fig.~\ref{fig:2}.  The output is a zero-mean stationary
Gaussian random process. As shown in Fig.~\ref{fig:2}\,(a), the
signal exhibits a slowly varying stochastic envelope, with alternating intervals of enhanced and suppressed amplitude. This behavior results from the interference among the Fourier components distributed within the finite axion linewidth: constructive interference produces large-amplitude oscillations, whereas destructive interference suppresses the field. Figure~\ref{fig:2}(b) shows a shorter time interval, where the rapid carrier oscillation at $f_a$ becomes clearly visible. The coexistence of this fast oscillation with a slowly varying random envelope reflects the separation between the axion Compton frequency and the much smaller
frequency scale associated with its spectral linewidth.
Over longer time scales, as shown in Fig.~\ref{fig:2}\, (c), the
waveform exhibits the noise-like behavior expected for a stochastic axion field, where the RMS coincides with the square root of the set power.

The temporal properties of the generated waveform can be quantitatively validated through its autocorrelation function. For a stationary process, the normalized autocorrelation is related to the target power spectral density by the Wiener--Khinchin theorem,
\begin{equation}
R_s(\tau)
=
\int_{-f_s/2}^{f_s/2}
S_a(f)e^{i2\pi f\tau}\,df .
\label{eq:autocorrelation}
\end{equation}
The autocorrelation obtained from the expected lineshape from Eq.~\ref{eq:lineshape}, can be directly compared with that obtained from the generated signal using the discrete PSD defined in Eq.~\ref{eq:discrete-PSD}.

\begin{figure}[htp]
    \centering
    \includegraphics[width=1\linewidth]{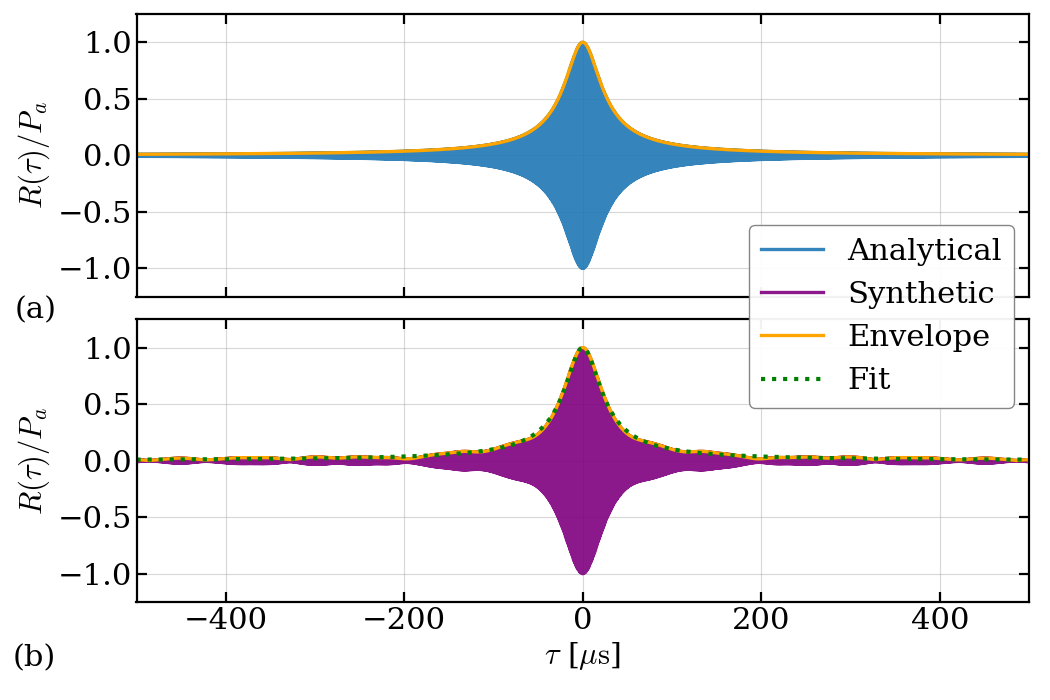}
    \caption{Comparison between the theoretical autocorrelation function and the one obtained from the synthetic data. (a)~The theoretical autocorrelation is defined with coherence time $\tau_a~\simeq~0.681\beta = 140.1~\mathrm{\mu \rm s}$. (b) The fit for the synthetic signal autocorrelation gives $\tau_a =141.9 \,\mu \rm s$. This result is consistent with the expected approximate value for the 6~GHz axion.}
    \label{fig:autocorrelation}
\end{figure}

To characterize the temporal coherence of the signal, we consider the decay of the autocorrelation envelope. While the autocorrelation may exhibit oscillations or statistical fluctuations, its slowly varying envelope is determined by the spectral linewidth $\Delta f_a$ and therefore provides the relevant timescale for coherence. Obtaining an analytical expression for the autocorrelation associated with the full laboratory-frame line shape is not straightforward. We therefore approximate the latter using the Galactic frame expression given in Eq.\ref{eq:unboosted-lineshape} with the width parameter rescaled as $\beta\rightarrow0.681\beta$ to account for the laboratory frame broadening. Under this approximation, the nor\-ma\-li\-zed autocorrelation function is 
\begin{equation}
    \frac{R_s(\tau)}{\Pa} = \Re \left\{\frac{e^{i2\pi f_a'\tau}}{\left(1-i\,2\pi\tau/\tau_a\right)^{3/2}} \right\},
\label{eq:analytical_autocorr}
\end{equation}
whose envelope is given by
\begin{equation}
    E(\tau)=\left[1+\left(\frac{2\pi \, \tau}{\tau_a}\right)^2\right]^{-3/4},
\label{eq:envelope_fit}
\end{equation}
where $\tau_a$ is the characteristic correlation timescale. Note that $\tau_a$ is the scale parameter of the adopted envelope and does not necessarily coincide with other operational definitions of coherence time. The characteristic timescale is set by the inverse spectral linewidth and therefore scales as $\tau_a\propto {1}/{\Delta f_a}$, where the numerical coefficient depends on the assumed lineshape and on the adopted definition of coherence time. For the parameters considered here, the expected envelope scale is
$\tau_a \simeq 0.681\beta = 140.1\,\mu\mathrm{s}$.
We estimate $\tau_a$ by fitting Eq.\ref{eq:envelope_fit} to the envelope of the synthetic signal autocorrelation function, as shown in Fig.\,\ref{fig:autocorrelation}.\\

\subsection{Spectral validation}

\begin{figure}[htp]
\includegraphics[width=1\linewidth]{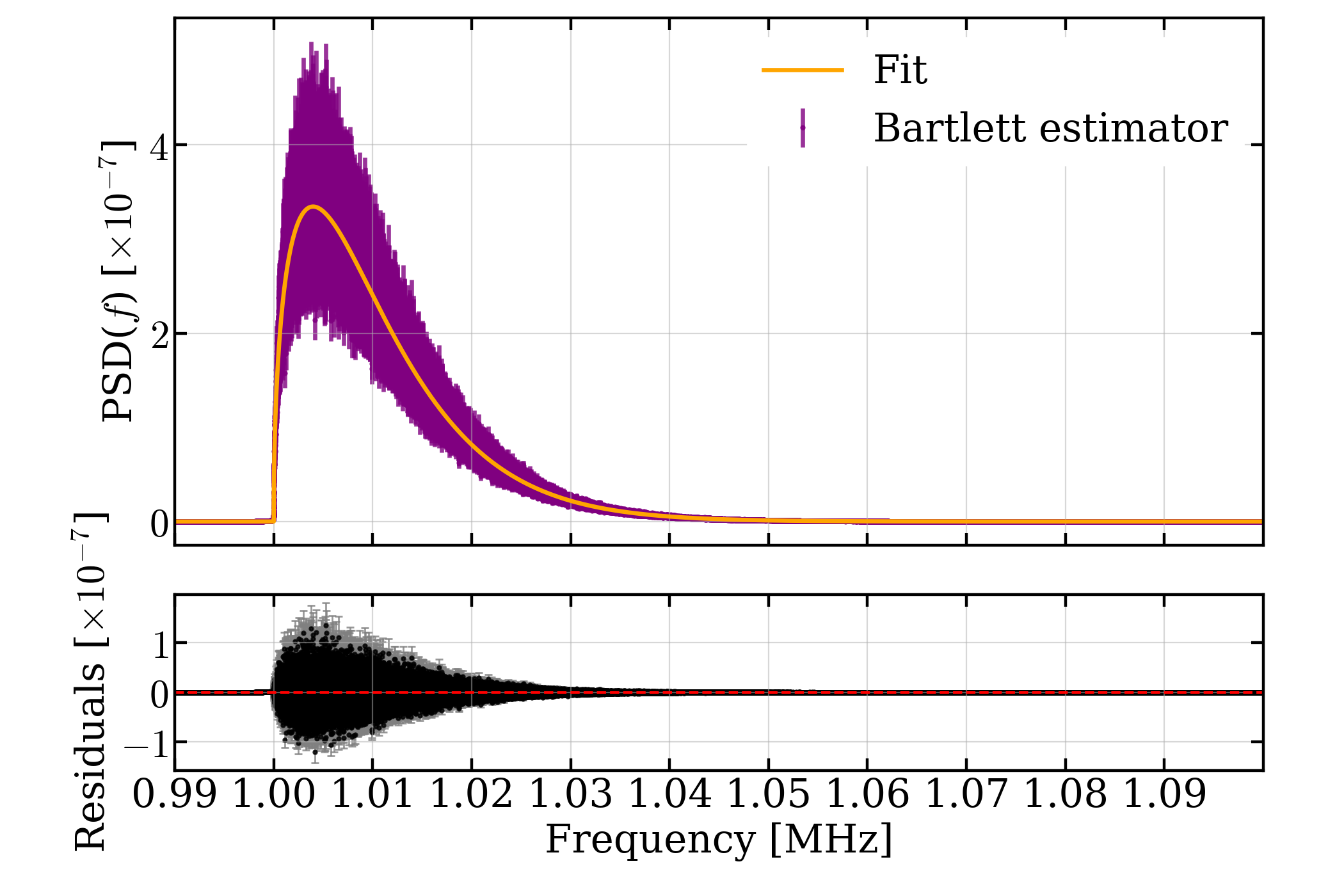}
\caption {Spectral validation of the generated synthetic axion signal.
The averaged power spectral density, obtained from 100 independent realizations, is shown together with a fit to the target axion line shape. The central frequency, $f_a' = 1~\mathrm{MHz}$, and the width parameter, $\beta = 2.054\times10^{-4}~\mathrm{Hz^{-1}}$, are free parameters in the fit, while $\Pa$ and $\alpha$ are fixed to their input values. The lower panel shows the residuals between the averaged spectrum and the best-fit model. The residuals exhibit larger fluctuations in regions of higher signal power, as expected for a stochastic signal. Only the positive-frequency component of the two-sided PSD is shown for clarity.
}
\label{fig:spectral}
\end{figure}

The spectral properties of the generated waveform are validated by
estimating its power spectral density and comparing it with the target axion line shape. Since the periodogram of a single realization exhibits large statistical fluctuations, the time series is divided into $N_B$ segments and the corresponding periodograms are averaged following
Welch's method~\cite{Welch:1967aa}. For statistically independent segments, the fractional standard deviation of the estimated power in each
frequency bin decreases approximately as $1/\sqrt{N_B}$.\\
It is important to emphasize that an individual realization is not
expected to reproduce the target line shape exactly. Rather, the
prescribed power spectral density describes the ensemble-averaged
spectrum of the stochastic axion field and is progressively recovered as the number of averaged periodograms increases.\\
To quantitatively validate the generated spectrum, the averaged PSD is fitted with the laboratory-frame axion line shape of
Eq.~\ref{eq:axion-lineshape}, as shown in Fig.\,\ref{fig:spectral}. In this parametrization, $f_a^{\prime}$
determines the signal frequency, while $\beta$ sets its characteristic spectral width through $\Delta f_a=1/\beta$. 
In the fit $f_a^{\prime}$ and $\beta$ are left as free parameters,
whereas $\alpha$ and the total signal power $P$ are fixed to
the values used in the signal generation. 
Both $f_a^{\prime}$ and $\beta$, as recovered from the fit, are consistent with the corresponding values used to generate the target spectrum. This demonstrates that the spectral-shaping procedure accurately reproduces both the position and the characteristic width of the prescribed axion line shape.

\subsection{Power normalization}

The power generated, measured in $\mathrm{V}_{\text{rms}}^2$, is evaluated from the variance in the time-domain,
\begin{equation}
P_{\mathrm{time}}=\frac{1}{N_{\mathrm{tot}}}
\sum_{n=0}^{N_{\mathrm{tot}}-1}s^2[n],
\label{eq:time-power}
\end{equation}
or by integrating the two-sided periodogram,
\begin{equation}
P_{\mathrm{freq}}
=
\sum_k S_M(f_k)\,\Delta f,
\label{eq:frequency-power}
\end{equation}
where $\Delta f$ is the width of the frequency bin.
Parseval's theorem requires the two estimates to agree, subject to the selected
FFT normalization and finite-sample fluctuations.

The injected digital power is then mapped to the physical power at the chosen injection plane using an independently measured attenuation or gain calibration. To express the resulting quantity in Watts [W], the mean-square voltage must first be divided by the load impedance:
\begin{equation}
P_{\mathrm{inj}}(f) =
\frac{P_{\mathrm{DAC}}(f)}{50\,\Omega} \,
10^{-A(f)/10} ,
\label{eq:physical-calibration}
\end{equation}
where $A(f)$ is the calibrated power attenuation between the digital-to-analog converter and the cavity input port.

\section{Experimental implementation and tests}

\begin{figure*}[htp]
    \includegraphics[width=0.49\textwidth]{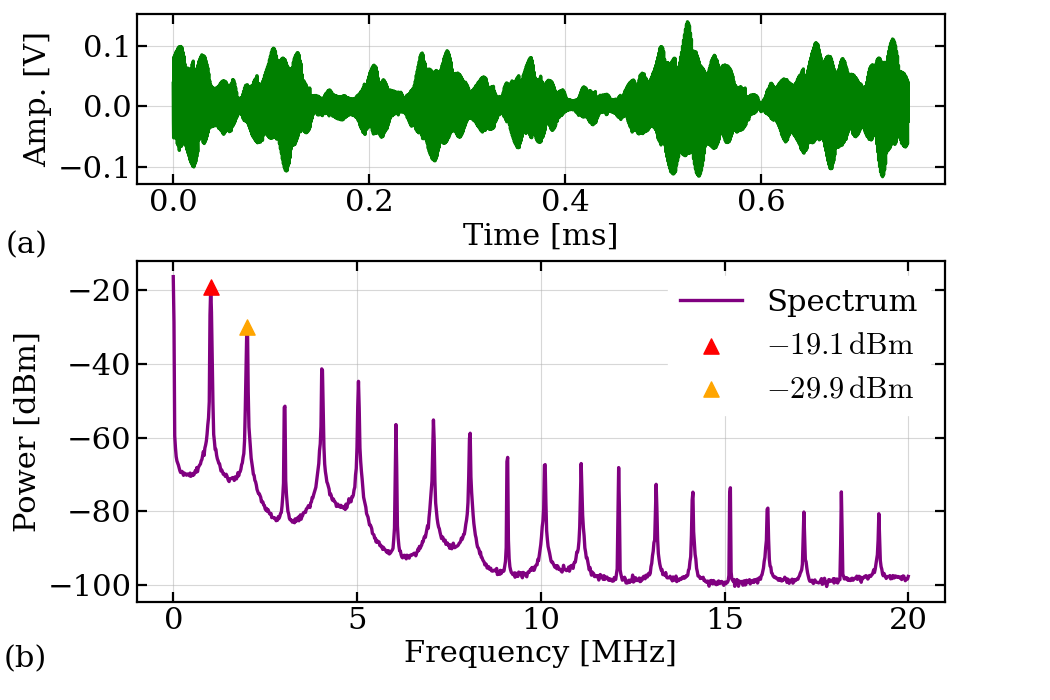}
    \includegraphics[width=0.49\textwidth]{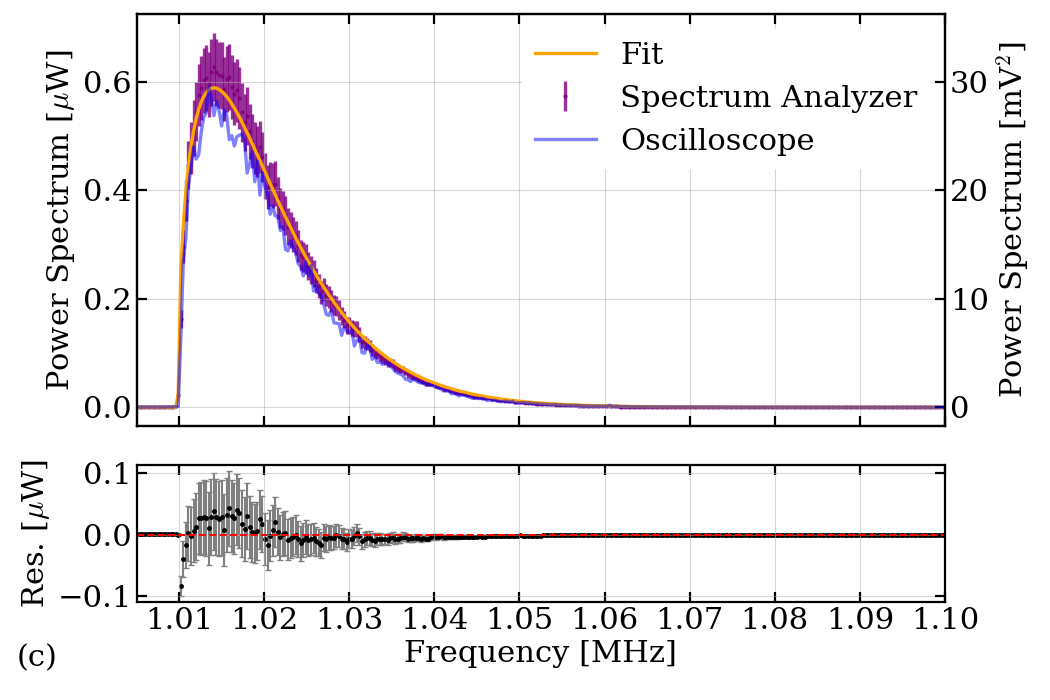}
    \caption {(a) Axion signal generated at $f_a' = 1\,$MHz with a power of 0.01\,V$^2$ as observed on the oscilloscope. (b) Averaged spectrum measured with a signal analyzer (SA). Spectral images whose amplitudes are modulated by a sinc-shaped envelope are visible. The first image to the right of the desired spectrum is suppressed by approximately 10\,dB and is indicated by the yellow marker. (c) Spectrum corresponding to the generated signal. The time-domain data acquired with the oscilloscope were Fourier-transformed and rebinned to match the frequency bins of the signal analyzer (blue points). The purple curve shows the spectrum averaged over 100 measurements at the SA, together with the corresponding fit (orange curve), giving $f_a = 1.009808$\,MHz, $\beta = 2.013\times10^{-4}~\mathrm{Hz^{-1}}$. The powers $P_{\rm SA} = 30.665~\mathrm{\mu W}$ at the SA is computed following Eq. \ref{eq:frequency-power} and it is a fixed parameter in the fit. Same holds for the oscilloscope power $P_o = 0.0014~\mathrm{V^2}$. The residuals of the fit are shown in the lower panel.
    }
    \label{fig:DAC}
\end{figure*}

\subsection{Digital-to-analog conversion}

The digitally generated synthetic axion signal must be converted into an analog waveform. 
In the present work, the conversion is performed using a Digital-to-Analog Converter (DAC), which combines a D/A converter with a sample-and-hold (S/H) circuit~\cite{Proakis:1996:DAC}. 

The D/A converter maps each numerical sample to a corresponding voltage level. The S/H circuit then keeps this voltage constant for one sampling interval $T_s=1/f_s$ until the next sample arrives. Consequently, the DAC output is a staircase-like approximation of the desired signal. An analog low-pass reconstruction filter is then used to smooth this waveform and suppress unwanted high-frequency components.
The spectral properties of this signal can be understood by separating the effects of sampling and holding. Sampling the signal at regular intervals produces shifted copies of its spectrum $S(f)$, centered at integer multiples of the sampling frequency $f_s$. 
Mathematically, the sample-and-hold output $S_{\rm SH}(f)$ can then be modeled as the sampled signal convolved with a rectangular pulse of duration $T_s$, which in the frequency domain yields:

\begin{equation}
S_{\mathrm{SH}}(f)
=
\left[
S(f) * \sum_{k=-\infty}^{\infty}
\delta(f-kf_s)
\right]
\cdot T_s
\frac{\sin(\pi f T_s)}{\pi f T_s}.
\label{eq:sample_and_hold}
\end{equation}
\noindent
This expression shows the finite holding time weights the signal spectral images with a sinc-shaped envelope.

To validate the proposed method, preliminary tests were performed using a DAC already available in the laboratory, with a maximum output rate of $3.3~\mathrm{MS/s}$. A sampling rate of $f_s = 3~\mathrm{MHz}$ is used to synthesize the axion signal at $f_a^{\prime}$, operating the device close to its maximum output rate. \\
As shown in Fig.\,\ref{fig:DAC}\,(a), the resulting spectral copies do not overlap with the axion baseband spectrum.

In addition, the signal parameters are successfully recovered, as shown in Fig.~\ref{fig:DAC}\,(b). The fit results are consistent with those expected for a \(6~\mathrm{GHz}\) axion signal centered at \(1~\mathrm{MHz}\). In addition, the fitted parameters agree with those obtained from the averaged periodograms of the corresponding digital signal before the DAC conversion, shown in Fig.~\ref{fig:spectral}.\\
Note that a significant fraction of the signal power is lost because of the attenuation introduced by the DAC low-pass filter. This effect has to be be taken into account in the calibration procedure.
 
\subsection{Upconverting the low frequency synthetic axion signal}

\begin{figure}[htp]
    \includegraphics[width=0.49\linewidth]{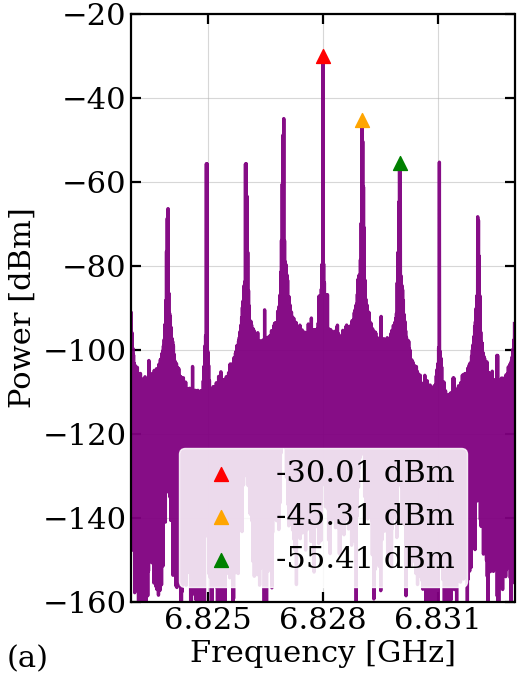}
    \includegraphics[width=0.49\linewidth]{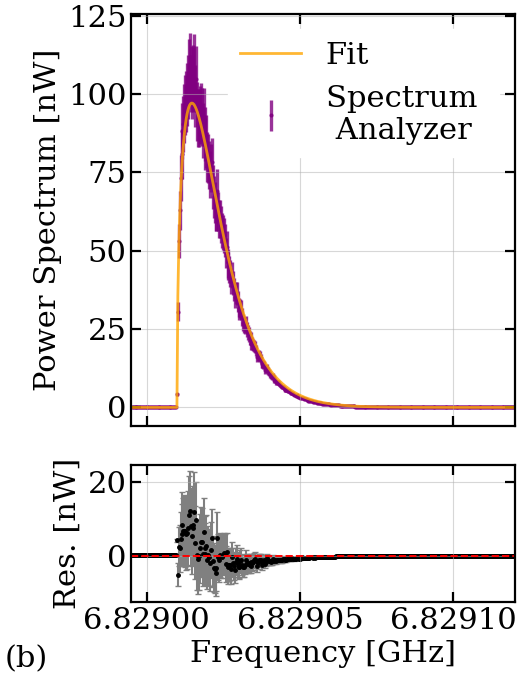}
    \caption{(a) Spectrum of the synthetic axion signal after upconversion to $f_a$. The carrier at $6.829~\mathrm{GHz}$ is indicated by the yellow marker,  together with the mixing products and the upconverted spectral replicas  generated by the DAC. 
    (b) Fit to the measured spectrum at $6.829~\mathrm{GHz}$, averaged over 100 acquisitions, with the corresponding residuals shown in the lower panel. The fitted parameters are $f_a = 6.82900994~\mathrm{GHz}$ and $\beta = 1.752 \times 10^{-4}~\mathrm{Hz}^{-1}$. The power $P_{\mathrm{SA}} = 5.803~\mu\mathrm{W}$ is computed as in Eq. \ref{eq:frequency-power}}
    \label{fig:mixer}
\end{figure}

The signal at the DAC output lies in the low-frequency range and must therefore be up-converted to the target microwave frequency without distorting its spectral profile.
This is accomplished with a double-balanced mixer~\cite{MiniCircuits:2020}, which combines a Local Oscillator signal at high-frequency $f_{\rm LO}$ with the low frequency axion frequency $f_a^{\prime}$ to obtain a synthetic axion signal centered at the target frequency
$f_{\mathrm{a}} = f_{\mathrm{LO}} \pm f_a^{\prime}$.

Although the axion signal $s(t)$ is not a pure sinusoid, its frequency-domain representation can be used to describe the mixing process. Multiplication by the LO carrier at frequency $f_{\mathrm{LO}}$ shifts the spectrum of the IF signal to the sum and difference frequencies:
\begin{equation}
V_{\mathrm{RF}}(f) \propto S(f-f_{\mathrm{LO}})+S(f+f_{\mathrm{LO}}).
\end{equation}
Thus, the spectrum of the low-frequency axion signal is translated around the LO frequency, with the upper sideband centered at $f_{\mathrm{LO}}+f_a^{\prime}$.

In the present setup, the synthetic axion signal generated by the DAC is centered at $f_a'=1$\,MHz and applied to the IF port of the mixer. The LO frequency is chosen such that the desired mixing sideband coincides with the resonance frequency \(f_c=6.829~\mathrm{GHz}\) of a cylindrical microwave cavity operating in the \(\mathrm{TM}_{010}\) mode. Selecting the upper sideband requires \(f_{\mathrm{LO}}=f_c-f_a'=6.828~\mathrm{GHz}\). The resulting up-converted signal is therefore centered on the cavity resonance, allowing the cavity to serve as a mock haloscope.
For the chosen axion frequency, the spectral-width parameter is $\beta=1.963\times10^{-4}\mathrm{Hz}^{-1}$.

The up-converted signal is then transmitted  through a microwave coaxial cable to a signal analyzer. Fig.~\ref{fig:mixer}\,(a) shows the measured spectrum after up-conversion. The LO carrier at $f_{\mathrm{LO}}=6.828~\mathrm{GHz}$ is clearly visible at the center, together with the two mixing products at $6.827~\mathrm{GHz}$ and $6.829~\mathrm{GHz}$. The spectral images generated by the DAC are also up-converted, with the first image attenuated by approximately $10~\mathrm{dB}$ relative to the desired signal. However, its power contribution at the cavity frequency is negligible, as it is $\sim 60$\,dB below the desired signal. 
As shown in Fig.~\ref{fig:mixer}\,(b), the total signal power decreases from $30.672~\mathrm{\mu W}$ to $5.802~\mu\mathrm{W}$ due to the conversion loss of the mixer.
The spectral width parameter $\beta$ and the axion frequency $f_a$ obtained from the same fitting procedure described above are consistent with the corresponding values measured at the DAC output prior to up-conversion. This agreement demonstrates that the mixer translates the signal to the target microwave frequency without altering its spectral profile.

\subsection{Injection of the synthetic axion signal into a test cavity}

\begin{figure}[h!]
    \centering
    \includegraphics[width=0.45\textwidth]{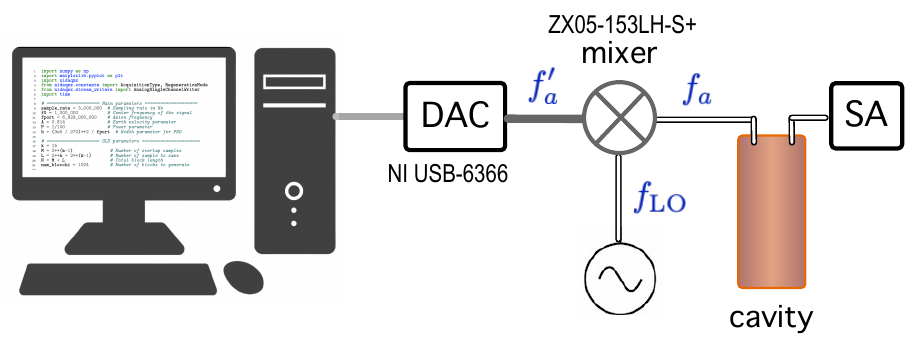}
    \caption{Experimental setup for the frequency up-conversion and microwave-cavity test. The DAC output is applied to the IF port of a double-balanced mixer, which translates the spectrum centered at $1~\mathrm{MHz}$ to the cavity resonance frequency, $f_c\simeq6.829~\mathrm{GHz}$, as measured at the signal analyzer (SA). The local oscillator is set to $f_{\mathrm{LO}}=6.828~\mathrm{GHz}$, such that the upper mixing sideband is centered on the cavity resonance.}
    \label{fig:diagram}
\end{figure}

The up-converted synthetic axion signal was subsequently injected into a room-temperature copper microwave cavity to investigate its transmission and filtering through a resonant structure representative of a cavity haloscope as illustrated in Fig.\,\ref{fig:diagram}. The cavity resonates at $f_c=6.829~\mathrm{GHz}$, and has a FWHM of about 430\,kHz. 

\begin{figure}[h!]
    \centering
    \includegraphics[width=0.49\linewidth]{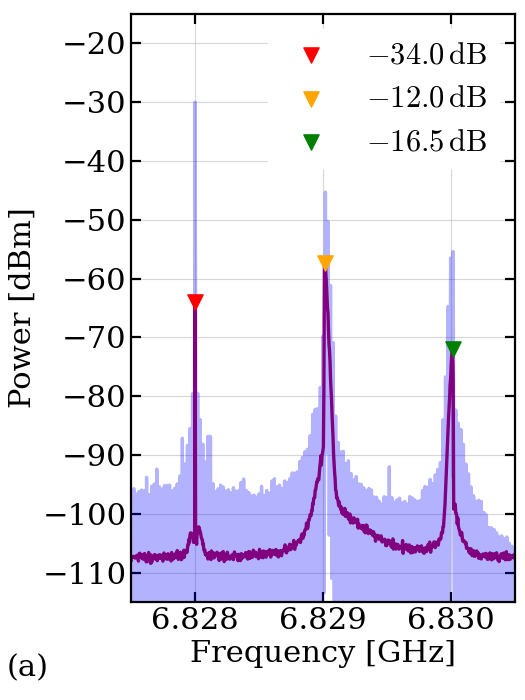}
    \includegraphics[width=0.49\linewidth]{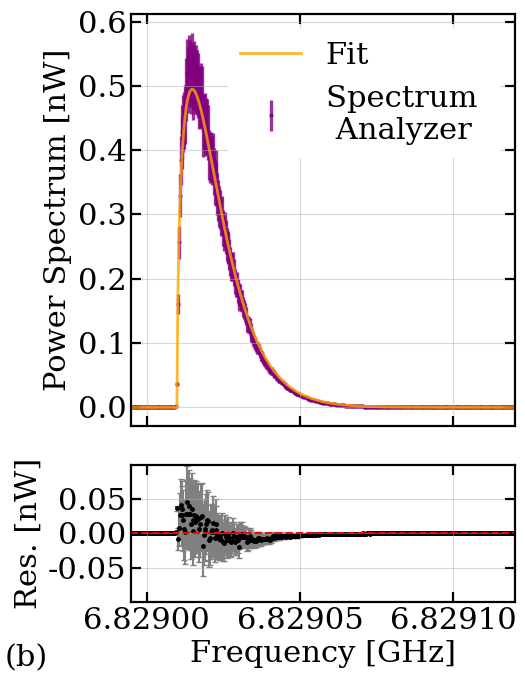}
    \caption{(a) Cavity-filtered spectrum measured with the signal analyzer and centered on the synthetic axion signal at $f=6.829~\mathrm{GHz}$, overlaid with the spectrum measured at the mixer output prior to injection into the cavity. The comparison shows the attenuation introduced by the cavity for the three spectral peaks. The lower-frequency peak is the mixing sideband, whereas the higher-frequency peak originates from a spectral image already present at the DAC output. (b) Enlarged spectrum of the synthetic axion signal after the cavity, together with the fit to the expected axion line shape. As in the previous analysis, the total signal power $P_{\mathrm{SA}}=30.165~\mathrm{nW}$ is held fixed and determined by summing the power over the frequency bins. The fit yields $f_a=6.82900989~\mathrm{GHz}$ and $\beta=1.73\times10^{-4}~\mathrm{Hz}^{-1}$.}
    \label{fig:axion_cavity}
\end{figure}

Because the nearest spectral images are separated from the desired signal by only $1\sim$\,{MHz}, the cavity used in this proof-of-principle setup is not sufficiently selective to suppress them completely, as shown in Fig.~\ref{fig:axion_cavity}\,(a). 

This technical limitation is not fundamental and can be mitigated by employing commercially available high-speed DACs or arbitrary-waveform generators with substantially higher sampling rates. The resulting increase in spectral separation would allow the unwanted images to be efficiently suppressed by the cavity or by an additional microwave filter. 

Additional rejection may be provided by the haloscope cavity itself. In particular, the high-$Q$ superconducting cavities increasingly employed or developed for axion searches \cite{Lee:2026HTS, Maiello:2026,DAhn:2022,Posen:2023} exhibit substantially narrower bandwidths and can therefore strongly attenuate sampling images located outside their resonance. 
The synthetic signal retains its spectral shape after the cavity (see Fig.\,\ref{fig:axion_cavity}\,(b)), with fitted parameters consistent with those obtained for the up-converted signal in Fig.~\ref{fig:mixer}. 
Overall, these results demonstrate that the generated signal can emulate the axion-induced excess power expected in a microwave cavity haloscope, provided that the sampling rate and filtering scheme are selected to ensure sufficient suppression of the spectral images.

\section{Conclusion}

We have described an FFT-based method for ge\-ne\-ra\-ting stochastic synthetic axion signals for microwave cavity haloscopes. Gaussian white noise is shaped by the square root of a target axion PSD, and overlap-save filtering permits continuous, arbitrarily long waveforms to be produced with controlled memory use. The resulting signal reproduces the desired ensemble spectrum, total power, random Fourier statistics, and finite-coherence time-domain behavior. The modular spectral model supports both the Standard Halo Model and more general velocity distributions.

Although developed for conventional haloscopes employing linear receivers, the proposed method is particularly attractive for next-generation quantum-enabled haloscopes based on microwave photon counters. In these experiments, realistic synthetic waveforms can be injected through a weakly coupled port to validate not only the receiver calibration but also the complete photon-counting protocol, including detector efficiency, dark-count subtraction, dead-time corrections, and statistical inference. The method therefore provides a unified calibration framework applicable to both linear-amplifier and photon-counting axion searches.

\begin{acknowledgments}
This work is supported by INFN (QUAX experiment), the Italian Ministry of University and Research (MUR) through the PRIN 2022 project
``Chasing dark matter with quantum technologies'' (Project No. 2022BP4H73) and the MUR Departments of Excellence grant 2023-2027 “Quantum Frontiers”.
\end{acknowledgments}

\section*{Data availability}
The data and code that support the findings of this study are available from the corresponding authors upon reasonable request.

\bibliography{biblio}

\end{document}